\documentclass[aps,prl,twocolumn,superscriptaddress,floatfix]{revtex4}
\usepackage{amsmath}
\usepackage{amsfonts}
\usepackage{amssymb}
\usepackage{graphicx}
\usepackage{cancel}
\usepackage[switch,columnwise]{lineno}
\usepackage{titlesec}
\usepackage{mathrsfs}

\makeatletter
\renewcommand{\section}{\@startsection{section}{1}{0mm}
  {-\baselineskip}{0.5\baselineskip}{\bf\leftline}}

\renewcommand{\subsection}{\@startsection{section}{1}{0mm}
  {-\baselineskip}{0.5\baselineskip}{\bf\leftline}}
\makeatother

\begin{document}

\title{Detecting dynamical quantum phase transition via out-of-time-order correlations in a solid-state quantum simulator}
\author{Bing Chen}%
\affiliation{School of Electronic Science and Applied Physics, Hefei University of Technology, Hefei, Anhui 230009, China}%
\affiliation{State Key Laboratory of Quantum Optics and Quantum Optics Devices, Shanxi University, Taiyuan 030006, China}%
\author{Xianfei Hou}%
\affiliation{School of Electronic Science and Applied Physics, Hefei University of Technology, Hefei, Anhui 230009, China}%
%\author{Dayou Yang}%
%\affiliation{Institute for Quantum Optics and Quantum Information, Austrian Academy of Sciences, Technikerstrasse 21a, 6020 Innsbruck, Austria}%
%\affiliation{Institute for Theoretical Physics, University of Innsbruck, Technikerstrasse 21a, 6020 Innsbruck, Austria}%
\author{Feifei Zhou}%
\affiliation{School of Electronic Science and Applied Physics, Hefei University of Technology, Hefei, Anhui 230009, China}%
\author{Peng Qian}%
\affiliation{School of Electronic Science and Applied Physics, Hefei University of Technology, Hefei, Anhui 230009, China}%
%\author{Yanqiang Guo}%
%\affiliation{Key Laboratory of Advanced Transducers and Intelligent Control System, Ministry of Education, College of Physics and Optoelectronics, Taiyuan University of Technology, Taiyuan, 030024 China}%
%\author{Georg Enzian}%
%\affiliation{Clarendon Laboratory, University of Oxford, Parks Road, Oxford, OX1 3PU, UK}%
\author{Heng Shen}%
\email{heng.shen@physics.ox.ac.uk}
\affiliation{Clarendon Laboratory, University of Oxford, Parks Road, Oxford, OX1 3PU, UK}%
\author{Nanyang Xu}%
\email{nyxu@hfut.edu.cn}
\affiliation{School of Electronic Science and Applied Physics, Hefei University of Technology, Hefei, Anhui 230009, China}%

%\author{Author1}%
%\affiliation{School of Electronic Science and Applied Physics, Hefei University of Technology, Hefei, Anhui 230009, China}%
%\author{...}%
%\affiliation{School of Electronic Science and Applied Physics, Hefei University of Technology, Hefei, Anhui 230009, China}%
%\author{Author2}%
%\affiliation{Clarendon Laboratory, University of Oxford, Parks Road, Oxford, OX1 3PU, UK}%

%\newcounter{mathematicapage}
\begin{abstract}
Quantum many-body system in equilibrium can be effectively characterized using the framework of quantum statistical mechanics. However, non-equilibrium behaviour of quantum many-body systems remains elusive, out of the range of such a well established framework. Experiments in quantum simulators are now opening up a route towards the generation of quantum states beyond this equilibrium paradigm. As an example in closed quantum many-body systems, dynamical quantum phase transitions behave as phase transitions in time with physical quantities becoming nonanalytic at critical times, extending important principles such as universality to the nonequilibrium realm. Here, in solid state quantum simulator we develop and experimentally demonstrate that out-of-time-order correlators, a central concept to quantify quantum information scrambling and quantum chaos, can be used to dynamically detect nonoequilibrium phase transitions in the transverse field Ising model. We also study the multiple quantum spectra, eventually observe the buildup of quantum correlation. Further applications of this protocol could enable studies other of exotic phenomena such as many-body localization, and tests of the holographic duality between quantum and gravitational systems.
\end{abstract}
\maketitle
%\textbf{Quantum many-body system in equilibrium can be effectively characterized using the framework of quantum statistical mechanics. However, non-equilibrium behaviour of larger scale systems remains elusive, out of the range of such conventional framework. Experiments in quantum simulators have now opened up a route towards the generation of quantum states beyond this equilibrium paradigm. As an example in closed quantum many-body systems, dynamical quantum phase transitions behave as phase transitions in time with physical quantities becoming nonanalytic at critical times, extending important principles such as universality to the nonequilibrium realm. Here, in the platform of solid state we develop and experimentally demonstrate that out-of-time-order correlators, a central concept to quantify quantum information scrambling and quantum chaos, can be used to dynamically detect nonequilibrium phase transitions in transverse field Ising model. We also study the multiple quantum spectra, attempting to observe the buildup of quantum correlation. Further applications of this protocol could enable studies other exotic phenomena such as many-body localization, and tests of the holographic duality between quantum and gravitational systems.}

Equilibrium properties of quantum matter can be effectively captured with the well-established quantum statistical mechanics. However, when closed quantum many-body systems are driven out of equilibrium, a lot of questions of how to understand the actual dynamics of quantum phase transition remain elusive since they are not accessible within thermodynamic description \cite{Eisert,Langen}. An exciting perspective arises from quantum simulators, which can mimic natural interacting quantum many-body systems with experimentally controlled quantum matter such as ultracold atoms in optical lattice and trapped ions. Such analogue system enables the investigation of exotic phenomena such as many-body localization \cite{Schreiber,Smith}, prethermalization \cite{Gring,Neyenhuis}, particle-antiparticle production in the lattice Schwinger model \cite{Martinez}, dynamical quantum phase transitions (DQPT) \cite{Jurcevic,Zhang,Flaschner} and discrete time crystal \cite{Zhang2,Choi}. 

In many of these phenomena, such as the celebrated logarithmic entanglement growth in many-body localization \cite{Znidaric,Bardarson,Altman}, the propagation of quantum information plays a central role, opening up new point of view and possibilities for probing out-of-equilibrium dynamics. To measure the propagation of information beyond quantum correlation spreading and characterize quantum scrambling through quantum many-body systems, the concept of out-of-time-order correlation (OTOC) is developed recently \cite{Swingle,Rey,Garttner,Li,Landsman}, leading to new insight into quantum chaos \cite{Maldacena,Hosur} and the black hole information problems \cite{Preskill,Shenker}.

Recent experimental progresses in measuring out-of-time-order correlation (OTOC) \cite{Garttner,Li,Landsman} deliver important new insight into a more thorough grasping of how such quantities characterized complex quantum system. For instance, OTOC can be used as entanglement witness via multiple quantum coherence \cite{Garttner2}, and in particular be used to dynamically detect equilibrium as well as nonequilibrium phase transitions \cite{Heyl,Dag}. Here, we emulate the dynamical quantum phase transition of quantum many-body system by a solid-state quantum simulator based on nitrogen-vacancy centre in diamond \cite{Chen}. Furthermore, measurement of OTOC is performed to quantify the buildup of quantum correlations and coherence, and remarkably to detect the dynamical phase transition.

\begin{figure*}
\centering
\includegraphics[width=0.9\textwidth]{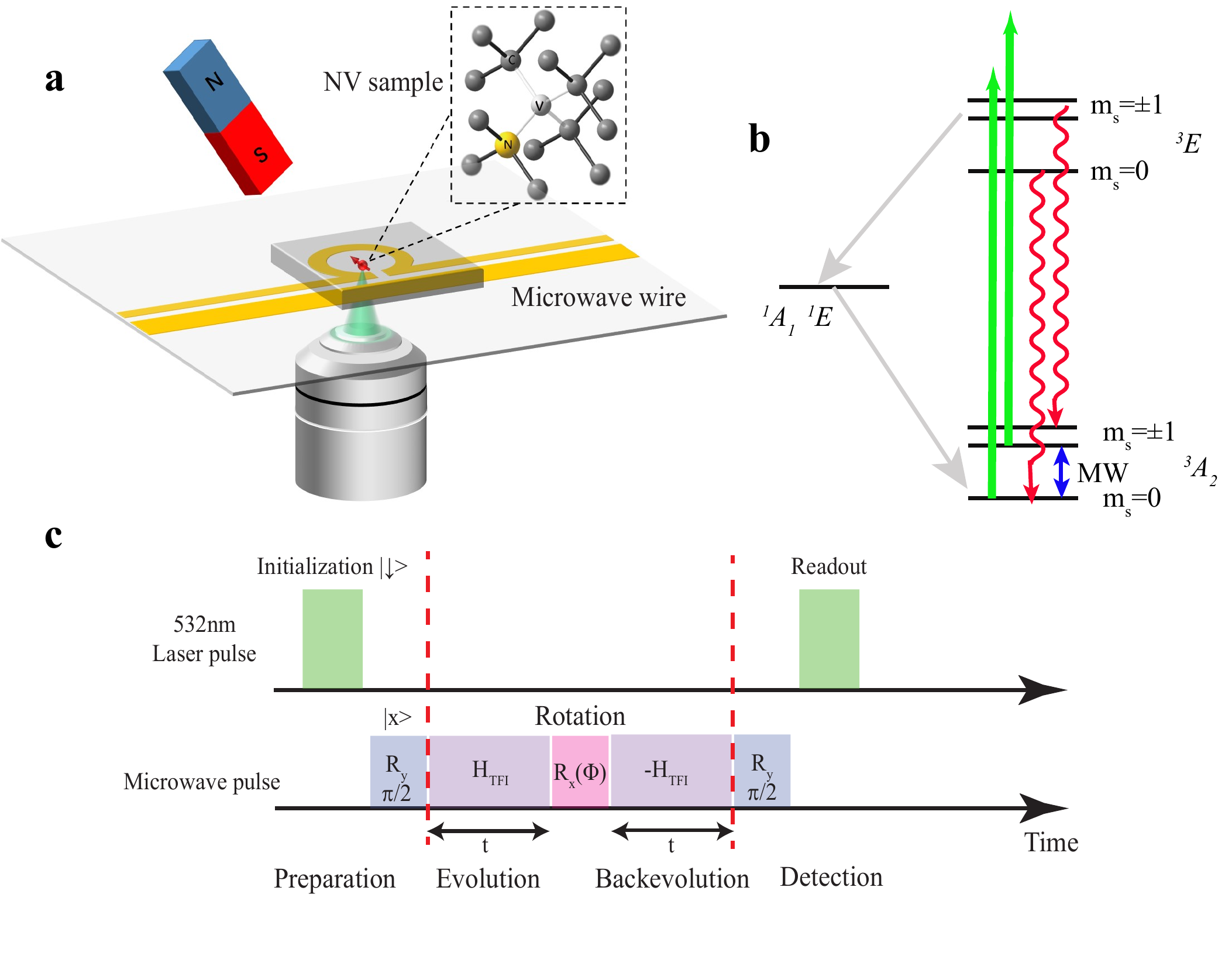}
\caption{\label{Fig:Setup} Measuring out-of-time-order correlation using time reversal in Nitrogen-vacancy centre. \textbf{a}. Illustration of experiment schematics and atomic structure of the Nitrogen-vacancy (NV) centre in diamond. \textbf{b}. Scheme of energy levels of the NV centre electron spin. Both its ground state (${}^3A_2$) and excited state (${}^3E$) are spin triplets. By applying a laser pulse of 532 nm wavelength with the assistance of intersystem crossing (ISC) transitions, the spin state can be polarized into $m_s=0$ in the ground state (${}^3A_2$). This process can be utilized to initialize and to read out the spin state of the NV centre. The fluorescence photons are detected by using the single photon counting module (SPCM). Additionally, a small permanent magnet in the vicinity of the diamond (magnetic field B $\approx$ 524 G) that is aligned parallel to the symmetry axis of the nitrogen vacancy centre splits the $m_s=\pm 1$ spin levels. With this magnetic field, the ${}^{14}$N nuclear spin of the NV centre can be also polarized with the laser pulse, which is enabled
by the level anti-crossing in its excited state. \textbf{c}. Laser and microwave pulse sequence for the measurement of OTOC. The $\pi/2$ rotation $\hat{R}_y$ about the y-axis prepares an initial state with spin pointing along x-axis $\left |x \right \rangle$. The state of interest $\hat{\rho}(t)$ is reached after the first evolution period. The rotation $\hat{R}_x(\phi)$ then imprints a phase $m\phi$ on each sector $\hat{\rho}_m$ of density matrix. Evolving backward and measuring the overlap with initial state as a function of $\phi$, the coherence $I_m$ and magnetization $A_m$ of $\hat{\rho}(t)$ are retrieved as the Fourier components of this signal.}
\end{figure*}

\begin{figure*}
\centering
\includegraphics[width=0.9\textwidth]{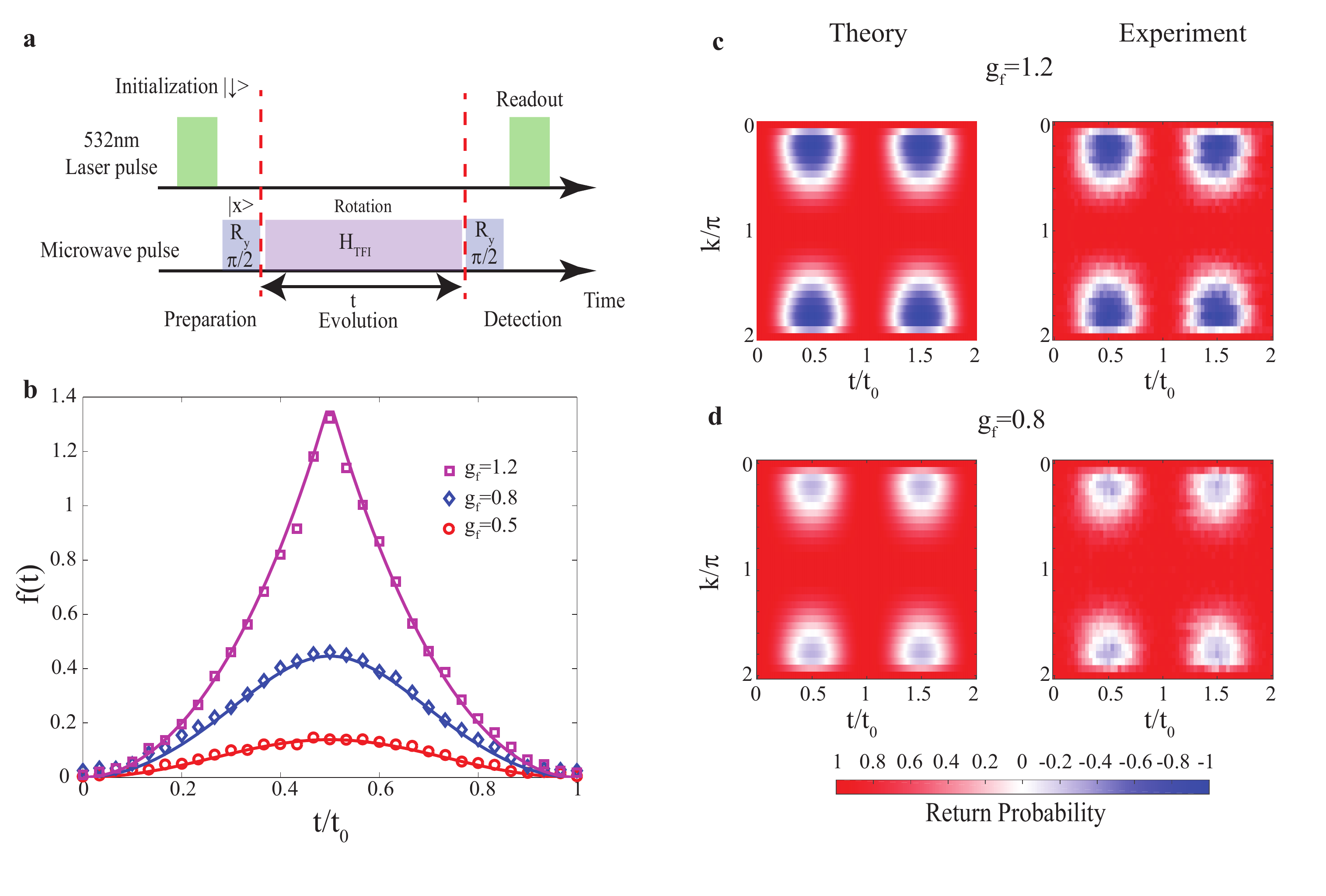}
\caption{\label{Fig2} Rate function dynamics after a quantum quench in an Ising model. \textbf{a}. Pulse sequence for DQPT. \textbf{b}. Rate function after a quantum quench of  $\hat{H}_{TFI}=-\sum_i^N(\sigma^x_i\sigma^x_{i+1}+g_f\sigma^z_i)$ with N=30.  The initial state is prepared at $\left | \phi_i \right \rangle=\left | x \right \rangle=(\left | \uparrow \right \rangle+\left | \downarrow \right \rangle)/\sqrt{2}$, as the ground state of $\hat{H}_0=-\sum_i^N\sigma^x_i\sigma^x_{i+1}$ (i.e. $g_i = 0$ in $\hat{H}_{TFI}$). $g_f$ is varied in the global quench with the values of 0.5, 0.8 and 1.2. Rate functions are shown in panel \textbf{b} with symbols (square, diamond, circle) and solid lines representing experimental and theoretical values. \textbf{c-d}. Return probabilities for $g_f=1.2$ and $g_f=0.8$, respectively. Theoretical results and experimental data are presented in the left and right panels, all sharing the same colorbar. For convince, $t$ is normalized by a period of time $t_0=\frac{\pi}{\left |\mathbf{d}_f(k)  \right |}$ (See Appendix).}
\end{figure*}

A very general setting for DQPT is the one emerging from a sudden global quench across an equilibrium quantum critical point \cite{Jurcevic,Zhang,Heyl2}, and DQPT manifests itself in discontinuous behaviour of the system at certain critical times. Here, we consider such a protocol. First, the state is initialized in the ground state of the initial Hamiltonian $\hat{H}_0=-\sum_i^N\sigma^x_i\sigma^x_{i+1}$ as $\left | \Phi \right \rangle_0=\left | + \right \rangle^{\bigotimes N}$ where $\left |+ \right \rangle=(\left | \uparrow \right \rangle+\left | \downarrow \right \rangle)/\sqrt{2}$ and N is the number of spins. At time t=0, the Hamiltonian is suddenly switched to $\hat{H}_{TFI}=-\sum_i^N(\sigma^x_i\sigma^x_{i+1}+g\sigma^z_i)$ and the system state evolves to $\left | \Phi (t) \right \rangle=e^{-i\hat{H}t}\left | \Phi \right \rangle_0$, realizing a quantum quench. Here $\mathbf{\sigma}=(\hat{\sigma}^x,  \hat{\sigma}^y, \hat{\sigma}^z)$ are Pauli spin operators. The rate function $f(t)$ as a function of return probability plays a role of a dynamical free energy, signalling the occurrence of DQPT. To explore DQPT of a spin-chain via a single solid-state qubit, $\hat{H}_{TFI}$ is written in momentum space as $\hat{H}_{TFI}=\sum_k\Psi_k^{\dagger} \hat{H}_k\Psi_k$ where $\Psi_k$ denotes a spinor with two elements vector composed of fermion operators (See Methods). The associated Bloch Hamiltonian is $\hat{H}_k=\mathbf{d}(k)\cdot\mathbf{\sigma}=[1-\cos⁡(k)g]\hat{\sigma}_x+\sin⁡(k)g\hat{\sigma}_y$ with $k$ the quasi-momentum. The bulk dynamics of the system can be solved since each k-component evolves independently, and after quench the state at each $k$ is given by$\left | \Phi (k,t) \right \rangle=e^{-i\hat{H}_kt}\left |\Phi (k,0) \right \rangle$. The rate function is defined as $f(t)=-1/N \sum_k \log(\left \langle \Phi (k,0) \left | e^{-i\hat{H}_kt} \right |\Phi (k,0)  \right \rangle)^2$, whose nonanalytic behaviour yields DQPT.

In the experiment we use a negatively charged NV centre in type-IIa, single-crystal synthetic diamond sample (Element Six) to simulate the quantum many-body dynamics in its quasi-momentum representation. As illustrated in Fig.\ref{Fig:Setup}, the NV centre has a spin triplet ground state. We encode $m_s=-1$ and $m_s=0$ in ${}^3A_2$ as spin up and down of the electron spin qubit. The state of the qubit can be manipulated with microwave pulses ($\omega_{MW}\approx2\pi\times$ 1400 MHz), while the spin level $m_s=+1$ remains idle due to large detuning. By applying a laser pulse of 532 nm wavelength with the assistance of intersystem crossing (ISC) transitions, the spin state can be polarized into $m_s=0$ in the ground state. This process can be utilized to initialize and read out the spin state of the NV centre. %The fluorescence photons are detected by using the single photon counting module (SPCM).
By using a permanent magnet a magnetic field about 524 G is applied along the NV axis, the nearby nuclear spins are polarized by optical pumping, improving the coherence time of the electron spin.

\begin{figure*}
\centering
\includegraphics[width=0.9\textwidth]{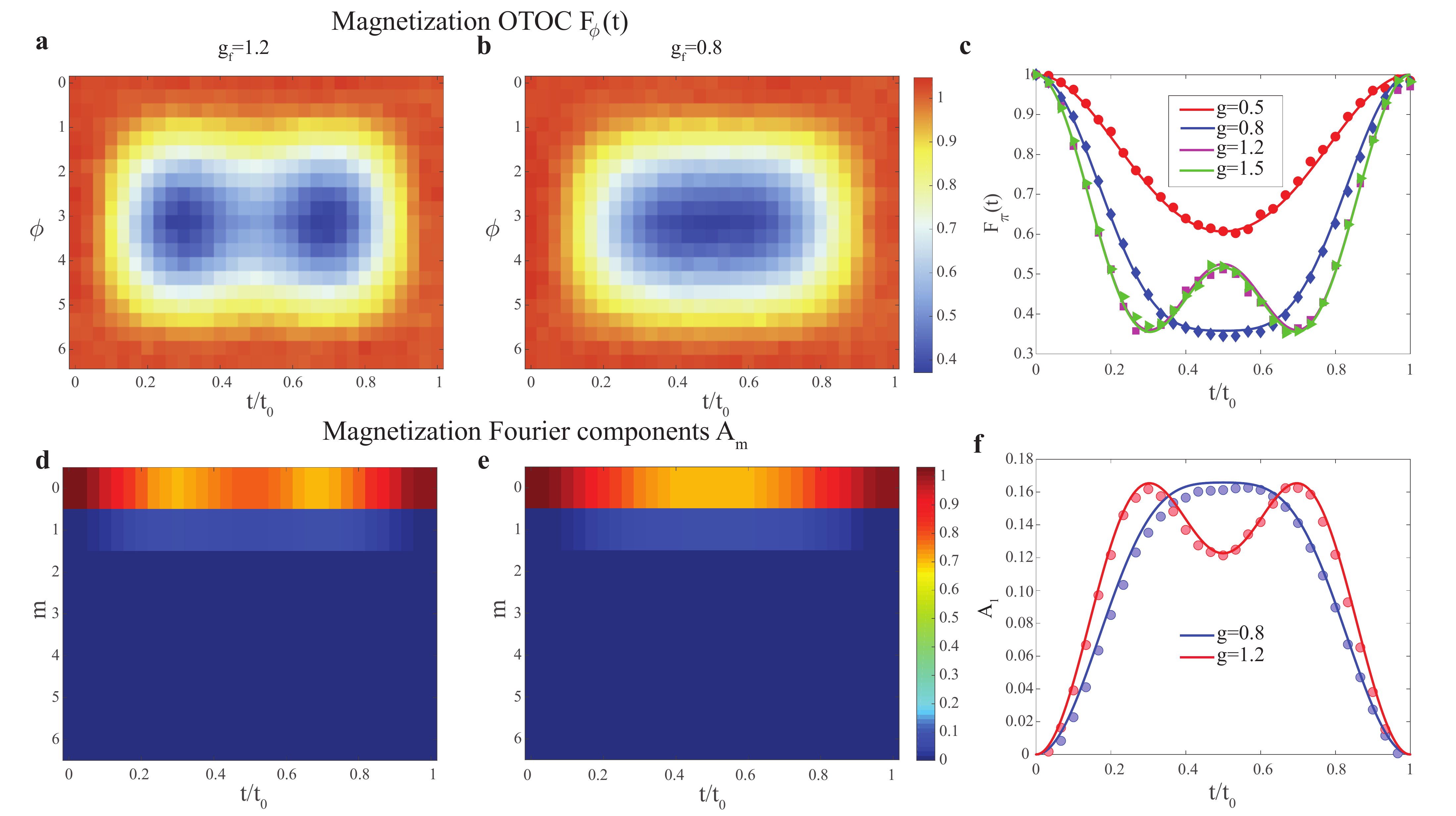}
\caption{\label{Fig3} Probing the DQPT through magnetization OTOC dynamics. The measured magnetization OTOC from time evolution under the $\hat{H}_{TFI}$ and rotations about the x-axis $0-2\pi$ with $g_f=1.2$ \textbf{a} and $g_f=0.8$ \textbf{b}. The associated Fourier components $A_m$ dynamics for $g_f=1.2$ \textbf{d} and $g_f=0.8$ \textbf{e}, respectively. \textbf{c}. The measured magnetization with fixed $\phi=\pi$ by varying $g_f$ (0.6, 0.8, 1.2, 1.5). \textbf{f}. Fourier components $A_1$ as a function of time where inversed double-well structure at the critical time $t_c$ is only shown in DQPT ($g_f>1$).}
\end{figure*}
Experimentally, evolution at different $k$ is performed in independent runs attributed to different rotation axes and speed (See Appendix). As sketched in Fig.\ref{Fig2}a, we prepare the initial state as $\left |+ \right \rangle=(\left | \uparrow \right \rangle+\left | \downarrow \right \rangle)/\sqrt{2}$, the ground state of $\hat{H}_0$, and then switch on the quench $\hat{H}_k$ by applying a resonant microwave pulse. The return probabilities are recorded by projecting the final states on x-basis. In the transverse field Ising model, the critical transverse field $g_c=1$ separates the paramagnetic phase ($g>1$) from ferromagnetic phase ($g<1$). Since state is initialized in ferromagnetic phase ($g_i=0$), DQPT only can occur only if $g_f>1$. It is demonstrated by the experimental data of the rate function for different $g_f$ values, as shown in Fig.\ref{Fig2}b, where the sharp peak with nonanalytic behaviour of rate function indicates the DQPT, being associated with dynamical Fisher zeros \cite{Vajna} (See Appendix).
%Fig.\ref{Fig2}c-d illustrates the emergence of Skyrmion lattice in momentum-time space for DQPT, indicated by the expectation value of the initial spin operator on the evolved state, $\hat{\sigma}_i(k,t)=2\left \langle \Phi (k,0) \left | e^{-i\hat{H}_kt} \right |\Phi (k,t)  \right \rangle^2-1$. When $g_i=0$ and $g_f>1$, Skyrmion lattice in momentum-time space appears at critical time, which proves that the non-trivial dynamical Chern number ensure the occurrence of DQPT.

Further signatures of the DQPT are observed by measure the OTOC, a quantity probing the spread of quantum information beyond quantum correlations. OTOC functions of particular interest is defined as Ref. \cite{Larkin,Garttner}, $F(t)=\left \langle \hat{W}^{\dagger}(t)\hat{V}^{\dagger} \hat{W}(t)\hat{V}\right \rangle$ where $\hat{W}(t)=e^{-i\hat{H}_{int}t}\hat{W}e^{i\hat{H}_{int}t}$ with $\hat{H}_{int}$ an interacting many-body Hamiltonian and $\hat{W}$ and $\hat{V}$ two commuting unitary operators. $\text{Re}[F(t)]=1-\left \langle \left | \left [ \hat{W}(t),\hat{V} \right ] \right |^2 \right \rangle/2$ captures the degree by which the initially commuting operators fail to commute at later times due to the many-body interactions $\hat{H}_{int}$, an operational definition of the scrambling rate. In such process the information initially encoded in the state spread over the other degrees of freedom of the system after the interactions, and cannot be retrieved by local operations and measurement.

We now outline the protocol to measure the OTOC as illustrated in Fig.\ref{Fig:Setup}c. In contrast to the pulse sequence shown in Fig.\ref{Fig2} a, we implement the many-body time reversal by inverting the sign of $\hat{H}_{TFI}$ which evolves again for time $t$ to the final state $\rho_f$ and ideally takes the system back to the initial state $\rho_0$. If a state rotation $\hat{R}_x(\phi)=e^{-i\hat{S}_x\phi}$, i.e. $\hat{W}(0)=\hat{R}_x(\phi)$, here about the x-axis with $\hat{S}_x=1/2\sum_i\hat{\sigma}_x^i$, is inserted between the two halves of the time evolution through a variable angle $\phi$, the dependence of the revival probability on this angle contains information about $\rho (t)$. At the end of sequence, two different observables can be measured, the collective magnetization along the x-direction, $\left \langle \hat{S}_x \right \rangle=tr[\hat{S}_x\hat{\rho}_f]$ and the fidelity $F_{\phi}(t)=tr[\hat{\rho}_0\hat{\rho}_f]$. In particular, the fidelity can be cast as an OTOC by setting $\hat{V}=\hat{\rho}_0$, corresponding to a many-body Loschmit echo. Measurement of the fidelity is also directly links to the so-called multiple quantum intensities $I_m$ by the Fourier transformation $F_{\phi}(t)=tr[\hat{\rho}_f\hat{\rho}_0]=tr[\hat{\rho}(t)\hat{\rho}_{\phi}(t)]=\sum_{m=-N}^{N}I_m(t)e^{im\phi}$ \cite{Garttner,Garttner2,Yao}.

Similarly, the dynamics of the Fourier amplitude $A_m$ of the magnetization $F_{\phi}(t)=\sum_{m=-N}^{N}A_m(t)e^{im\phi}$ quantifies the buildup of many-body correlations. However, it is much less sensitive to decoherence compared with the fidelity due to the nature of single-body observables. In Fig.3, we show the results of the magnetization OTOC measurement sequence and hence a buildup of Fourier amplitudes, $A_m$. Comparison of the data with $g_f=0.8$ to that with $g_f=1.2$ confirms that the appearance of double-well-like features in $\phi-t$ plane, signals the DQPT. Fig.\ref{Fig3}c shows the measured magnetization with fixed$\phi=\pi$ by varying $g_f$, in agreement with the conclusion above. The associated Fourier amplitude $A_m$ is extracted and illustrated in Fig.\ref{Fig3}d-e. Although the nearest-neighbour interaction limits the buildup of high order components, double-well feature still distinguishes the DQPT with non-DQPT. 

In summary, we demonstrate a new approach for investigating quantum many-body system out of equilibrium by a solid-state quantum simulator based on nitrogen-vacancy centre in diamond. The sharp peak with nonanalytic behaviour of rate function indicates the DQPT. Moreover, we measure the magnetization OTOC to quantify the buildup of quantum correlations and coherence, and in particular the intriguing feature arising from the dynamical phase transition is characterized to detect the occurrence of DQPT. Further applications of this protocol could enable studies of other exotic phenomena such as many-body localization, and tests of the holographic duality between quantum and gravitational systems.

\clearpage
%
%\section*{Methods}
%\appendix
\section*{Appendix}
\textbf{Spin chain model}
We start with a periodically driven spin chain with transverse-field Ising Hamiltonian as follows,
\begin{equation}\nonumber
\hat{H}_{TFI}=-\sum_n^N(\sigma^x_n\sigma^x_{n+1}+g\sigma^z_n),
\end{equation}
with g the transverse field strength. The first term describes the time-independent nearest-neighbour spin-spin coupling.

In order to obtain a single qubit Hamiltonian in momentum representation, we first apply the Jordan-Wigner transformation to fermionize $\hat{H}_{TFI}$, which is often used to solve 1D spin chains with non-local transformation \cite{Franchini}. For convenience, Pauli operators are expressed by spin raising and lowering operators as $\hat{\sigma}_n^{\pm}=(\hat{\sigma}_n^{x}\pm i\hat{\sigma}_n^{y})/2$. By applying Jordan-Wigner transformation, the original Hamiltonian is mapped to the free-fermion model with the definition of $\hat{\sigma}_n^+=e^{i\pi\sum_{j<n}f_j^{\dagger}f_j}f_n$, $\hat{\sigma}_n^-=f_n^{\dagger}e^{-i\pi\sum_{j<n}f_j^{\dagger}f_j}$ and $\hat{\sigma}_n^z=1-2f_n^{\dagger}f_n$. Here, the fermionic creation and annihilation operators $f_n^{\dagger}$ and $f_n$ satisfy the anti-commutation relations $\left \{ f_m,f_n \right \}=\left \{f_m^{\dagger},f_n^{\dagger}\right \}=0$ and $\left \{f_m,f_n^{\dagger}\right \}=\delta_{mn}$. The fermionized spin chain model reads as $\hat{H}=-\sum_n(f_n^{\dagger}f_{n+1}+f_n^{\dagger}f_{n+1}^{\dagger}+h.c.)+g(1-2f_n^{\dagger}f_n)$.

Next, by using Fourier transformation defined as $f_n=1/\sqrt{N}\sum_{k\in BZ}e^{ikn}f_k$ and $f_n^{\dagger}=1/\sqrt{N}\sum_{k\in BZ}e^{-ikn}f_k^{\dagger}$ with the quasi-momentum $k$ in the first Brillouin zone (BZ), the associated Hamiltonian can be written as $\hat{H}_{TFI}=\sum_k\Psi_k^{\dagger} [\sin(k)\sigma_y+(g-cosk)\sigma_z]\Psi_k$  in terms of the spinor basis $\Psi_k^{\dagger}=(f_k^{\dagger}, f_{-k})$. And we can denote $\hat{H}_k=\mathbf{d}(k)\cdot\mathbf{\sigma}$.

In our experiment, the initial state is prepared at $\left | \phi_i \right \rangle=\left | x \right \rangle=(\left | \uparrow \right \rangle+\left | \downarrow \right \rangle)/\sqrt{2}$, one eigenstate of $\hat{\sigma}_x$. Given $\left | \phi_i \right \rangle$ as the ground state of $\hat{H}_0$ in the protocol, a unitary transformation $\hat{U}$ should be used to meet $\hat{U}\hat{H}_0(k)\hat{U}^{\dagger}=\hat{\sigma}_x$, then the applied quench Hamiltonian $\hat{H}_{TFI}$ is rotated to $\hat{U}\hat{H}_{TFI}\hat{U}^{\dagger}$. In practice, unitary transformation $P$ and $S$ are introduced to diagonalize $\hat{\sigma}_x$ and $\hat{H}_0$ as $D=P^{-1}\sigma_xP$ and $\tilde{D}=S^{-1}H_0S$ respectively. By rewriting $\hat{\sigma}_x=PS^{-1}\hat{H}_0SP^{-1}$, one finds $U=PS^{-1}$.

The spin processes on the Bloch sphere with a period $\frac{\pi}{\left |\mathbf{d}_f(k)  \right |}$. In order to explore the full Brillouin zone $k$ should be varied from $0$ to $2\pi$ with $N+1$ steps (i.e. step size $2\pi/N$) and $N$ is equivalent number of spin.
For each $k$, a unitary rotation operation is applied with axis $\frac{\mathbf{d}_f(k)}{\left |\mathbf{d}_f(k)  \right |}$, i.e. $\Omega=C\left |\mathbf{d}_f(k)  \right |$ with Constant C from current experimental setting and $(\theta,\phi)=(\Omega t,\phi)$ with $\phi=\arcsin(\frac{\sin(k)g_f}{\sqrt{(1-\cos(k)g_f)^2+(\sin(k)g_f)^2}})$. Since we'd like to keep the normalized speed for all the $k$, pulse duration $T$ is varied from $0$ to $2\pi/\Omega$ with $N_T$ steps. In the experiment, we can fix it as $\sim100$ for instance.

\textbf{Critical time $t_c$}
In analogy to the Fisher zeros in the partition function, which trigger phase transition in equilibrium, dynamical Fisher zeros is introduced \cite{Vajna}. At dynamical Fisher zeros the Lochmidt amplitude
\begin{equation}\nonumber
\begin{aligned}
G(t)&=\left \langle \Phi_0\left | e^{-i\hat{H}_kt} \right |\Phi_0  \right \rangle\\
&=\prod_k [\cos (\left | d_f(k) \right |t)+id_i(k)\cdot d_f(k)\sin (\left | d_f(k) \right |t)]
\end{aligned}
\end{equation}
goes to zero. It requires the existence of the critical momentum $k^*$ at which the vector $d_f$ is perpendicular to $d_i$, i.e. 
$d_i(k)\cdot d_f(k)=0$. and DQPT occurs at $t_c=\frac{\pi}{\left | \mathbf{d}_f(k) \right |}(n+1/2)$, n=0, 1, 2....

Since we employ transverse field Ising model, one should satisfy the following condition,
\begin{equation}\nonumber
d_i(k)\cdot d_f(k)=\cos^2(k)+\sin^2(k)-(g_i+g_f)\cos (k)+g_ig_f=0
\end{equation}
which requires $k^*=\pm arccos\frac{1+g_ig_f}{g_i+g_f}$ on the condition of  $\left | \cos(k) \right |=\left | \frac{1+g_ig_f}{g_i+g_f} \right |<1$, causing to $sgn[(1-\left | g_i\right |)(1-\left | g_f\right |)]=-1$. This indicates DQPT occurs if and only if the initial and final Hamiltonian have to belong to different phases.

\textbf{Experiment setup}
The diamond used in this work is a $2 \text{mm}\times 2 \text{mm} \times500 \mu \text{m}$ type-IIa, single-crystal synthetic diamond sample (Element Six), grown using chemical vapor deposition (CVD) by Element Six, containing less than 5 ppb (often below 1 ppb) Nitrogen concentration and typically has less than 0.03 ppb NV concentration.

Single mode solid-state 532 nm laser is utilized to initialize and readout the electron spin state of the NV center. We can use an acoustic optical modulator (AOM) to control the laser and create the desired pulse, driven by an amplified signal from a home-built pulse generator. The fluorescence photons emitted from the NV center is collected by a 1.40 numerical aperture (NA) aspheric aplanatic oil condenser (Olympus), passed through a 600 nm longpass filter (Thorlabs) and a pinhole with a diameter of 50 $\mu\text{m}$, and detected by the single photon counting module (SPCM; Excelitas).

A small permanent magnet creates a bias magnetic field $B_0$ of 524 G along the NV axis, splitting $m_s=\pm1$ spin levels. With this magnetic field, the ${}^{14}$N nuclear spin of the NV center can be polarized based on optical pumping, improving the coherence time of electron spin. It is observed in the optically detected magnetic resonance (ODMR) spectra that the nuclear spin polarization is higher than 98\% \cite{Jacques}. 

\begin{figure*}\label{Fig4}
\centering
\includegraphics[width=0.9\textwidth]{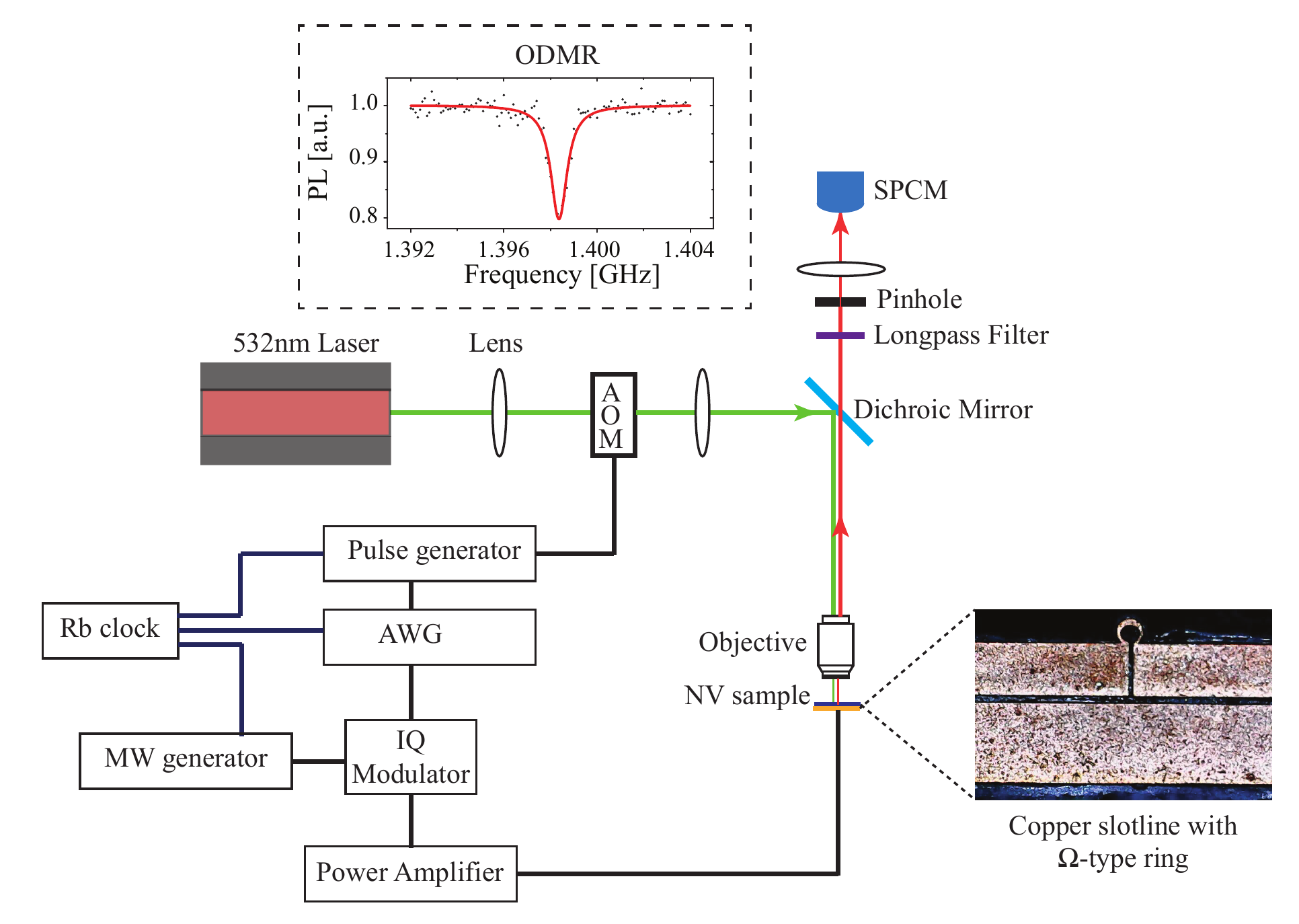}
\caption{Detailed schematic of the experimental setup. The solid colorful lines represent the path of the 532 nm laser and the fluorescence light. The solid black lines represent electrical connections.The above inset shows the ODMR signal which the microwave field frequency is about 1.4 GHz. The lower right corner inset is the copper coplanar waveguide which is the microwave field delivery setup. The impedance-matched copper coplanar waveguide with gap of 0.1 mm, a open-end $\Omega$-type ring in the middle and deposited on a coverslip. The outer diameter and inner diameter of the $\Omega$-type ring are 0.5 mm and 0.3 mm.}
\end{figure*}

Fig.\ref{Fig4} shows a schematic of the microwave (MW) setup. A commercial MW source (Rohde\&Schwarz) outputs a single frequency signal, mixed with the output from an arbitrary-waveform generator (AWG610; Tektronix; 2.6 GHz sampling rate) via IQ modulator to adjust the MW frequency and phase. Then the MW signal is amplified (Mini-Circuits ZHL-42W+) before delivery to an impedance-matched copper slotline with 0.1 mm gap, deposited on a coverslip, and finally coupled to the NV center. Note that pulse generator, MW source and AWG are all synchronized by locking to a 10 MHz reference rubidium clock.

%\section*{Data availability}
%The numerical data generated in this work is available from the authors upon resonable request.

\section*{Acknowledgements}
The authors are grateful to Dayou Yang, Philipp Hauke, Markus Heyl and Xiaojun Jia for fruitful discussions. 
This work is supported by the National Key R\&D Program of China (Grants No. 2018YFA0306600, and No. 2018YFF01012500), the National Natural Science Foundation of China (Grants No. 11604069 and No. 11904070), the Program of State Key Laboratory of Quantum Optics and Quantum Optics Devices (No. KF201802), the Fundamental Research Funds for the Central Universities, and the Natural Science Foundation of Anhui Province (Grant No. 1708085QA09). H. Shen acknowledges the financial support from the Royal Society Newton International Fellowship (NF170876) of UK.
%\section*{Author contributions}
%H.S. conceived the idea. N.X. and H.S. supervised the project. B.C. X.H. F.Z. and P.Q. performed the experiment. H.S. did the theoretical derivation and numerical calculations with contributions from all other authors. H.S. N.X. and B.C. wrote the manuscript with contributions from all other authors. All contributed to the discussion of the project and analysis of the experimental data.

%\section*{competing interests}
%The authors declare no competing interests.

\end{document}